\documentstyle [12pt,a4]{article}
\begin{document}
\begin{titlepage}
\newpage
\setcounter{page}{0}
\null
\vspace{2cm}

\begin{center}
{\Large {\bf On the Composite Fermion Approach
\vspace{0.5cm}

  in the FQHE}}

 \vspace{2cm}

{\large M.Eliashvili} $^{1,e,\dagger}$
  \vspace{0.5cm}

{\it Laboratoire de Physique Th\'eorique}\/ \rm
{\small E}N{\large S}{\Large L}{\large A}P{\small P} $^\ddag $ \\
{\it Chemin de Bellevue, B.P. 110, F-74941 Annecy-le-Vieux, Cedex, France.}

\end{center}

\vspace{2cm}

\centerline{\bf Abstract}

\indent

 \rm  FQHE is presented in the form of
non-unitary singular transformation, which relates the Laughlin
wave function (and its particle-hole conjugate)
to the composite quasi-particle incompressible
ground state.
\vspace{1cm}

\rightline{{\small E}N{\large S}{\Large L}{\large A}P{\small P}-A-478/94}

\rightline{July 1994}
\vspace*{\fill}

$\;$
\hrulefill\ $\; \; \; \; \; \; \; \; \; \; \; \; \; \; \; \; \; \; \; \;\;$
\hspace*{3.5cm}\\
\noindent
{\footnotesize $\;\ddag$ URA 14-36 du CNRS, assosi\'e  \`a l'E.N.S. de
Lyon, et au L.A.P.P. (IN2P3-CNRS) d'Annecy-le-Vieux}

 {\footnotesize $\;^1$ On leave of absence from Tbilisi Mathematical Institute,
  Tbilisi 380093, Georgia}

{\footnotsize $\;^e$ E-mail adress: merab@lapvax.in.2p3.fr}

{\footnotesize $\;^{\dagger}$adress after  September 1:
  Tbilisi Mathematical Institute,
  Tbilisi 380093, Georgia.}

\end{titlepage}

Practically all the essential information about the quantum Hall effect
can be encoded analytically in the form of the Laughlin wave function
\cite{laugh}
\begin{equation}
\Psi^e _p({\bf r}_1,...,{\bf r}_{N_e})=
\prod_{K< L}(z_K-z_L)^{2p+1}\exp \big[-\frac{eB}{4}
\sum_{I=1}^{N_e}|z_I|^{2}\big]
\end{equation}
which describes the incompressible ground state of $N_e$
spin-polarized planar electrons,
 with a complex coordinates $z_K=x_K+iy_K $,
moving in the orthogonal magnetic field $B=\varepsilon_{\alpha \beta}
A^{\beta}>0 (\alpha ,\beta =1,2).$
The integer $p$ is
related to the filling factor

\begin{equation}
\nu=\frac{N_e}{N_B} =\frac{1}{2p+1},
\end{equation}
where $N_B$ is the number of quantum states per Landau level.

When $N_e=N_B$, i.e. when $p=0$, (1) corresponds to the lowest Landau
 level completely filled
by the non-interacting electrons. For $N_e<N_B$ Laughlin function describes
the system of mutually interacting particles.

The notion of incompressibility recently was related to the
 the group of area preserving diffeomorphisms \cite{cap1},\cite{iso}, end
 at a quantum level two-dimensional
Hall fluid can be classified by the unitary irreducible representations
of the infinite dimensional algebra $W_{1+\infty}$, where (1) is a
highest weight vector \cite{cap2}.

The analytic and algebraic aspects of QHE are  supplemented  in the
 physically transparent way by the Jain's composite electron picture
 \cite{jain}.
In this picture FQHE is related  to the system
of non-interacting composite particles, consisting from electrons
bound to the  magnetic fluxes  $2p\phi_0$
 $(\phi_0=\frac {ch}{e}=\frac {2\pi}{e})$.

These fictitious magnetic fluxes can be associated with the singular
gauge potential ${\bf {a(r)}}$, such that the magnetic field
\begin{equation}
b({\bf r})=\varepsilon_{\alpha \beta }\partial_{\beta}a^{\beta}({\bf r})=
2p\phi_0 \varrho ({\bf r}),
\end{equation}
where $\varrho ({\bf r})$ is the particle density.

 Statistical gauge field
\begin{equation}
a^{\alpha}({\bf r})=-i \frac {p\phi_0}{\pi}
\partial_\alpha \int d{\bf r'}\arg \ln (z-z') \varrho ({\bf r'})
\end{equation}
 provides the required amount of
magnetic flux \cite{wil}, but the corresponding ground state turns
out to be compressible and can not be related to FQHE \cite{laugh1}.

As an alternative, there exists another gauge potential satisfying (3), the
 Knizhnik-Zamolodchikov connection \cite{kniz}  :
\begin{equation}
a^{\alpha}({\bf r})=-i\frac {p\phi_0}{\pi}
\partial_\alpha  \int d{\bf r'}\ln (z-z')\varrho ({\bf r'})
\end{equation}

The first-quantized form of
 covariant derivatives with this connection
\begin{equation}
D_I=\frac {\partial}{\partial z_I} -2p\sum_{J\not= J}\frac {1}{z_I-z_J},
\hspace{1cm}
\bar D_I=\frac {\partial}{\partial \bar z_I}
\end{equation}
has been used in  \cite{ver} to study the scattering problem for particles
obeying braid statistics, and in
 \cite{flo} to construct the Hamiltonian and $W_{1+\infty}$ algebra
generators for the Laughlin function (1) as a highest weight state.

It is not difficult to notice, that these covariant derivatives can
be presented in the form of the similarity transformation \cite{me}:
$$
D_I=S_p(z_1,\cdots,z_{N_e})\frac {\partial}{\partial z_I}
S^{-1}_p(z_1,\cdots,z_{N_e})
$$

$$
\bar D_I= S_p(z_1,\cdots,z_{N_e})\frac {\partial}{\partial \bar z_I}
 S^{-1}_p(z_1,\cdots,z_{N_e}),
$$
where
$$
  S_p(z_1,\cdots,z_N)=\prod_{K<L}(z_K-z_L)^{2p}
$$
is a singular non-unitary operator. This observation is helpfull in expressing
the wave function and quantum operators for the fractional value of
the filling factor as a similarity transformation of the
corresponding quantities for the non-interacting quasi-particle system.

Remark, that analogous transformations earlier have been introduced
  in order to get
a consistent  perturbative approach to a system of anyons
\cite{john},\cite{amel}
and as a mapping operator between the ground states with a different
filling factors \cite{amb},\cite{kar}.\footnote {Interesting to note,
 that non-unitary similarity transformations recently
have been considered  in the context of string quantum gravity,
indicating a deep conection between string theory  and
incompressible Hall fluid \cite{ellis}}

In this note we'll consider  the second-quantized version of
 this transformation.
Introduce the fermion Hamiltonian

\begin{equation}
H=\frac {1}{2m}\int d{\bf r}   (\partial_\alpha -ieA_\alpha({\bf r}))
\chi^\dagger ({\bf r})(\partial_\alpha +ieA_\alpha({\bf r}))\chi ({\bf r})
\end{equation}
 end define the transformed fields
$$
\psi ({\bf r})=S_p\chi ({\bf r})S^{-1}_p=
e^{2p\int d{\bf r}'\ln (z-z')\varrho ({\bf r}')}\chi ({\bf r}) ,
$$

$$
\psi^\star ({\bf r})=S_p\chi^\dagger ({\bf r}) S^{-1}_p=
\chi^\dagger ({\bf r})e^{-2p\int d{\bf r}'\ln (z-z')\varrho ({\bf r}')}
$$
 The operator $S_p$ can be presented as $S_p=e^{G_p}$,  where
the generator $G_p$ is a singular quadratic functional of the
density operator

$$
\varrho (x)=\psi^\star (x)\psi (x)=\chi^\dagger (x) \chi (x),
$$
and its explicit form is irrelevant for the present consideration.

  It is not difficult to show, that transformed fields obey
the normal Fermi statistics
$$
  \lbrace \psi ({\bf r}),\psi^\star ({\bf r}') \rbrace =
\delta ({\bf r}-{\bf r}')
$$
$$
  \lbrace \psi ({\bf r}),\psi ({\bf r}') \rbrace =
     \lbrace \psi^\star ({\bf r}),\psi^\star ({\bf r}') \rbrace =0
$$

Following \cite{jain} we'll interpret $\psi ({\bf r})$ as the electron
field, and associate $\chi ({\bf r})$ with the composite particles
carrying the even ($2p$) number of magnetic flux quanta.
In terms of the electron field the
 Hamiltonian (7) reads as follows:
$$
H= \frac {1}{2m}
\int d{\bf r}( \partial_\alpha \psi^\star
-ie\psi^\star A_\alpha -ie\psi^\star a_\alpha)
( \partial_\alpha \psi
 +ieA_\alpha \psi  +ie a_\alpha \psi) ,
$$
where
 $$
a_{\alpha}({\bf r})=i\frac {2p}{e} \partial_\alpha \int d{\bf r'}
\ln (z-z')\varrho({\bf r'}).
$$

In the Fock space where act the operators $\chi ({\bf r})$ and
$\chi^\dagger ({\bf r}) $,
the Hermitian conjugation is defined by the scalar product
\begin{equation}
\langle \Phi_1 |\hat O^\dagger | \Phi_2 \rangle=
\overline{\langle \Phi_2 |\hat O | \Phi_1 \rangle}
\end{equation}
It is evident, that the operators $\psi ({\bf r})$ and $\psi^\star ({\bf r})$
 are not Hermitian
conjugate in the sence of (8). Introduce the metric
operator $\hat \eta$ and define a new scalar product
$$
\langle \Phi_1 |\hat \eta \psi^\star ({\bf r}) | \Phi_2 \rangle =
\overline {\langle \Phi_2 |\hat \eta \psi ({\bf r})| \Phi_1  \rangle}
$$
This operator will be given by
$$
\hat \eta =(S^{-1}_p)^\dagger\cdot S^{-1}_p,
$$
and its matrix element in the $N$-particle subspace
 coincides with the inegration measure
introduced in \cite{ver},\cite{flo}.

The coordinate representation  bra- and ket vectors are
generated by the action of the physical electron
field   $\psi ({\bf r})$ on the vacuum (which is not changed under
the action of $S_p$),
 end are defined by
the expressions
$$
\langle z_1,...,z_N|=\langle 0|\psi (1) \psi (2) \cdot \cdot
\cdot \psi (N)=
$$
$$
 \langle 0|e^{2p  \sum_{K=1}^N \int dr  \ln (z-z_K)\varrho (r)}
\prod _{J< L}e^{2p\ln (z_J-z_L)} \chi (1) \chi (2)\cdot \cdot
\cdot \chi (N)=
$$
$$
\prod_{K<L}(z_K-z_L)^{2p}\langle 0|\chi (1)\cdots \chi (N)
$$
end

$$
| z_1,...,z_N \rangle=\psi^\star (N) \cdots \psi^\star(2)
\cdot \psi^\star (1)|0\rangle=
$$
$$
 \chi^\dagger (N)\cdots \chi^\dagger (2)
 \chi^\dagger (1)\prod _{J< L}e^{-2p\ln (z_J-z_L)}
 e^{-2p\int dr [\sum_{K=1}^N \ln (z-z_K)]\varrho ({\bf r}')} |0\rangle =
$$
$$
\prod_{K<L}(z_K-z_L)^{-2p}\chi^\dagger(N)\cdots \chi^\dagger (1)|0\rangle
$$

The quasi-particle field satisfy the Schr$\ddot {\rm o}$dinger equation for
the fermion in the uniform magnetic field.
Expand $\chi ({\bf r}) $  into modes
$$
\chi ({\bf r})=\chi_0 ({\bf r})+\tilde \chi ({\bf r}) ,
$$
where
$$
\chi_0 ({\bf r})=\sum_{j=0}^{N_B-1} f_j u_j({\bf r})
$$
contains only the lowest Landau level wave functions. For the disk geometry
and symmetric gauge ${\bf A}=\frac {B}{2}(-x,y)$
they are angular momentum eigenfunctions
\begin{equation}
u_j({\bf r})\sim  z^je^{-\frac {eB}{4} |z|^2}
\end{equation}
 The Fock space operators
satisfy usual fermionic anticommutation relations
$$
\lbrace f_j,f^+_l \rbrace =\delta _{jl}
$$

The modes corresponding to the lowest Landau level satisfy condition
$$
(\frac {\partial}{\partial \bar z}+\frac {eB}{4} z)\chi_0 ({\bf r})=0
$$
and the similarity transformation does not cause the level mixing
$$
 (\frac {\partial}{\partial \bar z}+\frac {eB}{4} z)
S_p\chi_0 ({\bf r})S_p^{-1}=0
$$

The ground state of Hamiltonian $H$ is extremly degenerate. All the
$N$-particle states of the form
$$
|N\rangle =f^+_{j_1}f^+_{j_2}\cdots f^+_{j_N}|0\rangle
$$
have the same energy.
One can select the particular ground state applying the Bogoliubov's
concept of quasi-averages \cite{bogo}. Following this method
 modify the Hamiltonian by the infinitesimal
perturbation, which lifts the degeneracy and find the unique ground
state.
After performing the necessary calculations
 and taking the thermodynamic limit, the perturbation is switched
off, leaving the results marked by this particular ground state.

In the case under consideration such a degeneracy lifting
naturally arises due to the external confining potential which
keeps particles together.  This circumstance selects as the ground state
the state with a minimal angular momentum.
 It has been shown, that  the transition to the states with a
higher angular momentum costs energy \cite{cap3}, promoting the
incompressible  state of noninteracting quasi-particles
  as a unique candidate for the ground state.

Consider the  Hamiltonian eigenstate
$$
|\Omega;N \rangle =\prod_{j=0}^{N-1}f^+_j |0\rangle
$$

 Case $N=N_B$ corresponds to the complete filling of the lowest Landau level.
The Laughlin ground state is given by
$$
\langle z_1,...,z_{N_e}|\Omega ;N_e \rangle = \prod_{K<  L}(z_K-z_L)^{2p}
\langle 0|\chi (1)\cdot \cdot
\cdot \chi (N_e)|\Omega ;N_e \rangle ,
$$
where the last factor
$$
\langle 0|\chi (1)\cdots \chi (N_e)f^+_{0}\cdots f^+_{N_e-1} |0\rangle =
\prod_{1\leq K<L\leq N_e}(z_K-z_L)e^{-\frac{eB_{ext}}
{4}
\sum_{I=1}^{N_e}|z_I|^{2}}
$$
 is the Slater determinant of one-particle states (9).

Cosequently, the Laughlin function can be defined as a similarity
transformation of incompressible state of $N_e$ non-intracting
composite particles. Since  $N_e\not= N_B$, this state does not
 correspond to the $\nu =1$ ground state, which for the given value of external
magnetic field would be  given by $|\Omega;N_B>$.
Remind, that in the original Jain's picture it is assumed that
the composite particles
exactly fill up the lowest Landau level, or in the other words FQHE is
interpreted as IQHE of bound states.

 The spectrum
generating quantum operators are related by the $S_p$ transformation to
 the corresponding
quantities of the non-interacting quasi-particle theory.
In particular, $W_{1+\infty}$ is generated
by the operators

$$
V^i_n= -\int d{\bf r}\chi^\dagger({\bf r})(B^\dagger_0)^{n+i}
(B_0)^i\chi ({\bf r})=
-\int d{\bf r}\psi^\star({\bf r})(B^\dagger_p)^{n+i}(B_p)^i\psi ({\bf r})
$$
where
$$
B_p=\frac {\partial}{\partial z}+ \frac {eB}{4} \bar z -
2p\int d {\bf r'}\frac {\varrho ({\bf r'})}{z-z'}
$$

$$
B^\dagger_p=-\frac {\partial}{\partial \bar z} +\frac {eB}{4}z
$$

Unlike the conformal field theory,   the
 generators $V_n^i$ are bounded from below $(n+i\geq 0)$, i.e. they form
so called "wedge" $W_\Lambda=\lbrace V_n^i,|n|>i\rbrace$, plus
the positive modes $n>i$  \cite{cap1} .

The same time the operators

$$
W^i_n=-\int d{\bf r}\psi^\star({\bf r})\exp\lbrace nB^\dagger_p\rbrace
(B_p)^i\psi ({\bf r})
$$
 satisfy the $W$-algebra commutation relations for any integer
 $n\in Z\hspace{-0,6em}Z ,i\geq 0$, i.e. they correspond to the
full $W_{1+\infty}$.

When $p\not= 0$, only $N_e=\frac{N_B}{1+2p}$ states in the lowest
Landau level are occupied.  The remaining
$N_h=N_B-N_e=N_B(1-\frac{1}{1+2p})$ states are empty, i.e. they are
described by the hole wave function \cite{girv}
$$
 \Psi^h _p(N_e+1,...,N_e+N_h)=
\langle \Omega;N_B|S_p^{-1}\psi^{\star}(N_e+N_h)\cdots \psi^{\star}(N_e+1)
|\Omega;N_e\rangle =
$$
$$
\int\cdots\int \prod_{K=1}^{N_h} [d{\bf r}_K]
\langle \Omega;N_B|S_p^{-1}
 \psi^{\star}(N_e+N_h) \cdots
\psi^{\star}(N_e+1)\psi^{\star}(N_e)\cdots \psi^{\star}(1)|0\rangle \times
$$
$$
\langle 0|\psi (1)\cdots\psi (N_e)
|\Omega;N_e\rangle =
$$
$$
\int\cdots\int \prod_{K=1}^{N_e} [d{\bf r}_K]
\overline {\Psi^e_0(1,\cdots,N_B)}\times \Psi^e_p(1,\cdots,N_e)
$$

The same wave function reappears  considering the particle-hole
conjugate system, or equivalently the system of electrons
in the magnetic field $-B<0$. Consider Hamiltonian

$$
H_c=\frac {1}{2m}\int d{\bf r}   (\partial_\alpha +ieA_\alpha)
\chi_c^\dagger ({\bf r})(\partial_\alpha -ieA_\alpha)\chi_c ({\bf r})
$$
where $\chi_c$ is a composite hole field.

The physical holes are introduced by the transformations
$$
\psi_c ({\bf r})=\bar S_p\chi_c ({\bf r})\bar S^{-1}_p=
e^{2p\int d{\bf r}'\ln (\bar z-\bar z')\varrho_c ({\bf r}')}\chi_c ({\bf r}),
$$

$$
\psi_c^\star ({\bf r})=\bar S_p\chi_c^\dagger ({\bf r}) \bar S^{-1}_p=
\chi_c^\dagger ({\bf r})e^{-2p\int d{\bf r}'\ln (\bar z-\bar z')
\varrho_c ({\bf r}')}
$$

$$
(\varrho_c (x)=\psi_c^\star (x)\psi_c (x)=\chi_c^\dagger (x) \chi_c (x))
$$
 Express the quasi-hole Hamiltonian  in terms of physical fields
$$
H_c= \frac {1}{2m}
\int d{\bf r}( \partial_\alpha \psi_c^\star
+ie\psi_c^\star A_\alpha +ie\psi_c^\star \bar a_\alpha)
( \partial_\alpha \psi_c
 -ieA_\alpha \psi  -ie \bar a_\alpha  \psi_c )
$$
where
 $$
\bar a_{\alpha}({\bf r})=-i\frac {2p}{e} \partial_\alpha \int d{\bf r'}
\ln (\bar z-\bar z')\varrho_c({\bf r'})
$$
is the charge-conjugate to the connection (5).

The charge-conjugate operators can be used to construct the state
vectors for the filling factors others then (2).

Expand  $\chi_c ({\bf r})$ into the modes
$$
\chi_c ({\bf r})=\sum_{j=0}^{N_B-1}f_{cj}u_{cj}({\bf r})+\tilde
\chi_c({\bf r}), \hspace {15mm} ( u_{cj}({\bf r})=\bar u_j({\bf r})  )
$$
and define vacuum  $|0_c\rangle$ (completely filled Fermi-Dirac sea):

$$
\chi_c({\bf r})|0_c\rangle =0
$$
Introduce the incompressible quasi-hole state
$$
|\Omega_c;N \rangle =\prod_{j=0}^{N-1}f_{cj}^+|0_c\rangle,
$$
and the state representing $N_B$ holes in the lowest Landau level
\begin{equation}
\bar S_p|\Omega_c;N_B \rangle
\end{equation}
Remark, that (10) is not an eigenvector of the Hamiltonian $H_c$.
Then
$$
\langle \Omega_c;N_B|\bar S^{-1}_p\psi^\star_c(1)\cdots \psi^\star_c(N_e)
|\Omega_c;N_h \rangle=
$$
\begin{equation}
=\int\cdots\int \prod_{K=N_e+1}^{N_B} [d{\bf r}_K]
\langle \Omega_c;N_B|\bar S^{-1}_p
\psi^\star_c (1)\cdots
\psi^\star_c(N_B)|0_c\rangle \times
\end{equation}
$$
\langle 0_c|\psi_c (N_B)\cdots \psi_c (N_e+1)|\Omega_c;N_h \rangle=
$$
$$
=\int \prod_{K=N_e+1}^{N_B} [d{\bf r}_K]
\overline {\Psi^e_p(N_e+1,\cdots,N_B)} \times \Psi^e_0(1,\cdots,N_B)
$$
will be the
 electron wave function  for the filling factor
$$
\nu=1-\frac{1}{1+2p}
$$
So, the "chiral" partners of  (6)
$$
\ D_I=\frac {\partial}{\partial  z_I},
\hspace{1cm}
\bar D_I=\frac {\partial}{\partial \bar z_I}
 -2p\sum_{J\not= J}\frac {1}{\bar z_I-\bar z_J}
$$
are engaged in the charge conjugate sector of the theory.

Considering the incompressible ground state of $m$ different species
 of noninteracting quasi-electrons $(N_e=\sum_{A=1}^m N_A)$,
one can relate it  to the wave function corresponding to the
filling factor  $\nu=\frac{m}{2mp+1}$ by the similarity transformation

\begin{equation}
S_{p,m}=\prod_{A<B} \prod_{I<J}(z^A_I-z^B_J)^{K_{A,B}}\prod_A
\prod_{I<J}(z^A_I-z^A_J)^{K_{A,A}-1}
\end{equation}
where the $m\times m$ matrix $K_{AB}$ is defined by \cite{fro}

$$
  K=\left|
\begin{array}{cccc}
2p+1 & 2p & ... & 2p \\
2p   &2p+1 &... & 2p \\
 .   &  .  & .  & . \\
2p  &...   & 2p &2p+1
\end{array}
\right|
$$

\vspace{1.5cm}
{\large {\bf Acknowlegments}}\/\rm \vspace{0.5cm}

I would like to thank all the members of  ENSLAPP for the kind
hospitality.
I'm particularly thankful to P.Sorba for bringing the considered problems
to my attention and permanent support. It is a pleasure to
thank N.Mavromatos for the interest and useful discussion.

\vspace{0,5cm}

\end{document}